\documentclass[%
 reprint, superscriptaddress,
 amsmath,amssymb,
 aps,
nolongbibliography]{revtex4-2}

\usepackage{graphicx}
\usepackage{dcolumn}
\usepackage{bm}
\usepackage{siunitx}
\usepackage{amsmath}
\usepackage{bbold}

\usepackage[normalem]{ulem}
\usepackage[hidelinks]{hyperref}

\usepackage{xcolor}
\definecolor{amber}{rgb}{1,0.49,0}

\usepackage{verbatim}

\usepackage{lineno}

\begin{document}
\title{Resolving the Nature of the Lowest-Frequency Raman Mode of Liquid Water}

\author{Florian Pabst
}
\email{fpabst@sissa.it}
\affiliation{%
 SISSA 
--
Scuola Internazionale Superiore di Studi Avanzati, 
34136
Trieste, Italy
}
\author{Harald Forbert}
\affiliation{%
 Center for Solvation Science ZEMOS, Ruhr-Universit{\"a}t Bochum, 44780 Bochum, Germany
}
\author{Dominik Marx}
\affiliation{%
 Lehrstuhl f{\"u}r Theoretische Chemie, Ruhr-Universit{\"a}t Bochum,  
44780
Bochum, Germany
}

\date{\today}

\begin{abstract}
The lowest-frequency Raman mode of water, observed through depolarized light
scattering or optical Kerr effect techniques, is routinely used to track dynamic
changes in water molecules near ions or biomolecules.
Yet, the microscopic origin of this mode and its relation to dielectric relaxation
still remains debated for pure water with conflicting interpretations attributing
it to either translational or rotational molecular motions.
In this study, we compute the low-frequency Raman spectrum in the GHz to THz range
using ab initio simulations, achieving excellent agreement with experimental data. 
Detailed decomposition analysis reveals that the
rotational and translational contributions are equally important,
while strong negative orientational cross-correlations as well as internal field
effects significantly modify the rotational component, making it distinct from
expectations inferred from dielectric spectroscopy.
\end{abstract}

\maketitle


Nearly 100 years ago, shortly after his Nobel Prize-winning discovery of
inelastically scattered light from liquids, Raman and his coworker reported a
"nebulosity or wings" in the scattered spectrum close to the incident
radiation \cite{raman1928molecular,raman1928rotation}.
Observing that these wings were unpolarized and their intensity scaled with the
anisotropy of the molecules, they attributed this part of the spectrum to the
rotational motions of molecules in the liquid.
However, in the late 1960s, it was discovered that even light scattered by
isotropic molecules is not fully polarized as previously
expected \cite{mctague1969intermolecular}.
This phenomenon was attributed to the dipole-induced-dipole (DID)
mechanism \cite{kielich1967role}: 
The incident light induces a dipole moment in a molecule, which subsequently
induces a dipole moment in neighboring molecules, resulting in an effective
anisotropy.
Since the DID effect depends on the intermolecular distances, it is commonly
associated with translational motions.

Despite these findings, the lack of a universally accepted theory of light
scattering in liquids has left unresolved whether rotational or translational
motions dominate the low-frequency Raman spectrum.
Typically, this part of the spectrum is attributed to rotational motions if
the molecule is significantly anisotropic and to translational motions
otherwise \cite{cummins1996origin}.
For water, however, both interpretations coexist in the literature.
Optical Kerr effect studies, for instance, attribute the lowest-frequency Raman
mode to rotational motions \cite{winkler2000ultrafast}, translational
motions \cite{gonzalez2023lifting}, or both \cite{palese1994femtosecond}.
Additionally, its appearance in the depolarized light scattering spectrum, at
a frequency similar to that of the secondary mode in the dielectric spectrum,
is often regarded as evidence of rotational
motions \cite{fukasawa2005relation,hansen2016identification}.
This interpretation stems from the fact that dielectric spectroscopy primarily
monitors the rotational motions of molecular electric dipole moments.
In this context, the dominant dielectric peak of neat water at approximately
\SI{18}{GHz} is attributed to long-lived hydrogen-bond-mediated orientational
cross-correlations between neighboring molecules, which are absent in the
Raman spectrum.

Experimental evidence has demonstrated that these cross-correlations explain
the discrepancy between dielectric and light scattering spectra for various
liquids \cite{pabst2021generic,pabst2020dipole}.
For water, in stark contrast, the situation is more complex. 
A recent ab initio simulation study \cite{holzl2021dielectric} revealed that
the single-molecule rotational mode in the dielectric spectrum occurs at
\SI{30}{GHz}, which cannot be reconciled with the lowest-frequency Raman mode
at around \SI{200}{GHz}, even accounting for a theoretically expected factor
of three between the two techniques \cite{diezemann1999revisiting}.
This suggests that instead of rotational motions, the lowest-frequency Raman
mode might be dominated by the DID effect, i.e., translational motions.
Indeed, early 
force field
simulations seemed to support this
hypothesis \cite{frattini1987induced}.
However, these simulations were based on anisotropy values obtained in vacuum,
which are now known to be underestimated \cite{lupi2012dynamics}.

Given these longstanding controversies, our goal is to clarify the yet unresolved
microscopic origin of the low-frequency Raman spectrum of pure water using
state-of-the-art ab initio molecular dynamics (AIMD) \cite{marxhuttermonograph}
simulations. 
Given that the electronic structure is accessible allows for the calculation of
spectra in close agreement with experiments as has been demonstrated for the
dielectric spectrum of water \cite{holzl2021dielectric,pabst2025salt}. 
 
We compute the Raman spectrum in the susceptibility representation
via \cite{berne2000dynamic,ramirez2004quantum}
\begin{equation}
    \chi''(\omega) = \frac{\omega}{6 k_{\rm B}TV} \int\limits_0^{\infty}  e^{-i \omega t} 
\sum_{\kappa\neq \lambda} \left< { A}_{\kappa\lambda}^{\rm eff}(0) { A}_{\kappa\lambda}^{\rm eff}(t)\right>_{\rm cl}{\rm d}t ,
    \label{eq:chi}
\end{equation}
where $T$ is the temperature, $V$ the volume and ${\bm A}^{\rm eff}$ is the
total effective polarizability tensor of the simulation box;
note that what is sometimes called 
``harmonic
quantum correction factor'' is
embodied in this expression following the Kubo formalism~\cite{ramirez2004quantum}
as shown in the Supplemental Material. 
Since we are interested in the depolarized spectrum, only the off-diagonal
elements ${\kappa \neq\lambda}$ are used. 
Here, ${\bm A}^{\rm eff}$ is calculated from the effective molecular
polarizability tensors as
${\bm A}^{\rm eff} = \sum_{i=1}^N {\bm \alpha}_i^{\rm eff}$, 
which in turn are obtained from the dipole moments induced by an applied external
electric field $E_0$,
\begin{equation}
    {\bm \mu}_i^{\rm ind} =  {\bm \mu}_i^E - {\bm \mu}_i^0 = {\bm \alpha}_i^{\rm eff} \bm E_0
\enspace , 
\end{equation}
as explained in the Supplemental Material. 

\begin{figure}[htb!]
    \centering
    \includegraphics[width=1\linewidth]{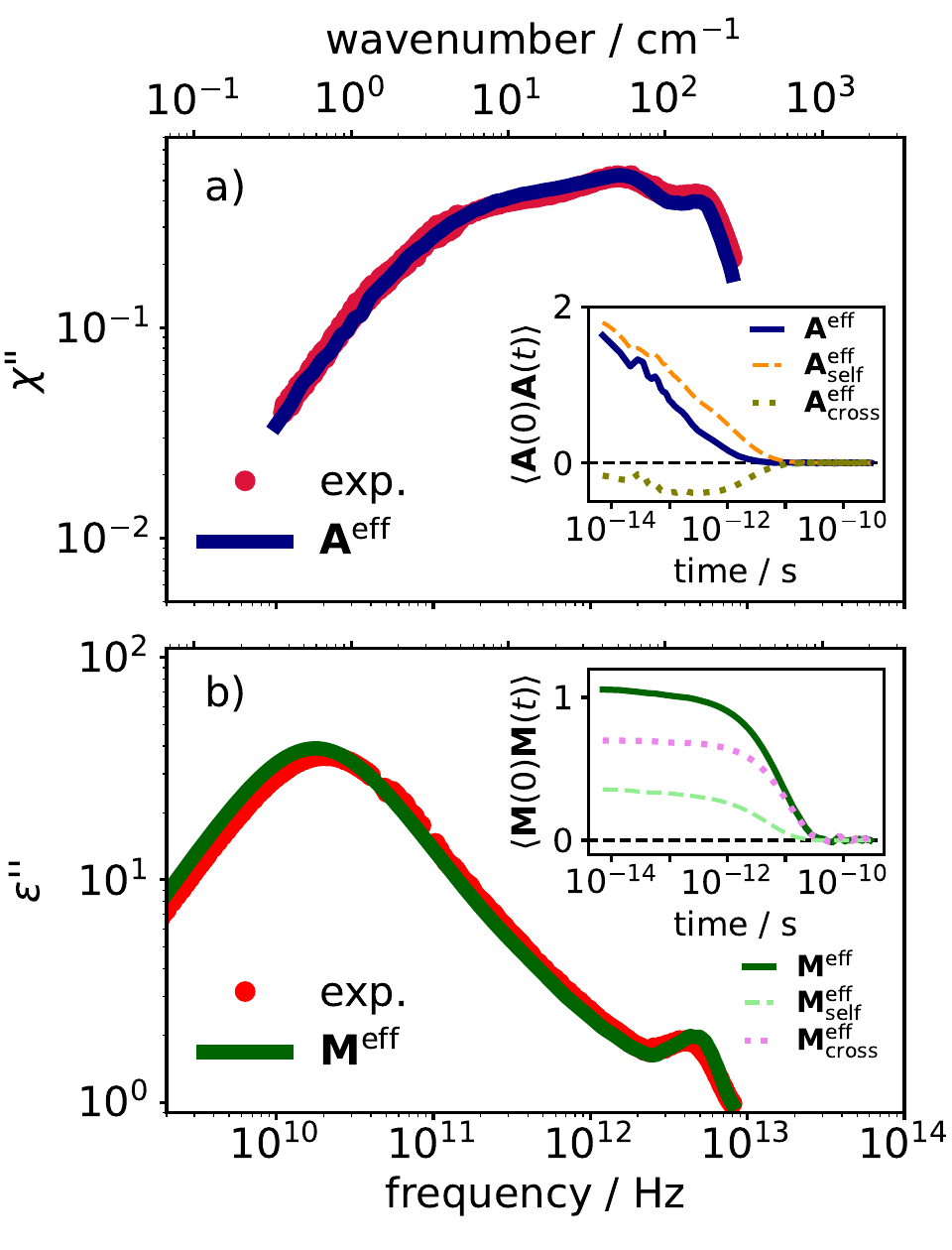}
    \caption{a) Comparison of the experimental depolarized light scattering spectrum 
\cite{hansen2016identification} with the one calculated from the AIMD simulations using Eq.~(\ref{eq:chi}). 
Inset: Total, self- and cross-correlation part of the polarizability correlation function underlying the calculated spectrum in the main panel. 
b) Same for the dielectric loss spectrum adapted from Ref.~\citenum{holzl2021dielectric}
and compared to the experimental dielectric spectrum~\cite{lunkenheimer2017electromagnetic}.}
    \label{fig:1}
\end{figure}

The dipole moments are calculated from maximally localized Wannier
functions~\cite{marzari2012maximally,marxhuttermonograph} computed from available
AIMD trajectories~\cite{holzl2021dielectric}.
We refer to these polarizability tensors as \emph{effective}, following earlier
terminology~\cite{wan2013raman}, because they represent the response observed
externally under the influence of an external field.
As such, they incorporate both the intrinsic polarizabilities of the molecules
and the effects of internal fields \cite{aubret2019understanding} as well as  
the DID interactions.
Consequently, this Raman spectrum is directly comparable to experimental
depolarized light scattering data~\cite{hansen2016identification} as shown
in Fig.~\ref{fig:1}a) without adjustments.
It is important to note that, in such experiments, the absolute intensity of the
spectrum is typically not recorded, although it can be measured
in principle \cite{pabst2022intensity}. 
As a result, the experimental spectrum has been vertically shifted on the y-axis
to align with the calculated spectrum.
The agreement between the two is excellent which provides together with the
similarly accurate dielectric loss in panel~b) a solid basis for comprehensive
spectral dissections and assignment as follows.

In the inset of Fig.~\ref{fig:1}a), the total, self- and cross-correlation part
of the polarizability correlation function
\begin{multline}
    \left< {\bm A}(0) {\bm A}(t)\right> = \\ \underbrace{\left< \sum\limits_{i=1}^{N} \bm\alpha_i(0) \bm\alpha_i(t) \right>}_{\rm self} + \underbrace{\left< \sum\limits_{i=1}^{N}  \bm\alpha_i(0)  \sum\limits_{j\neq i}^{N} \bm\alpha_j(t) \right>}_{\rm cross}
\end{multline}
are depicted. 
While the self part is slightly more intense than the total, the cross-correlation
is found to be negative throughout.
This is surprising since for the dielectric  spectrum the cross-correlation part
is roughly a factor of three more intense than the self part, thus dominating the
total dielectric loss, see inset of Fig.~\ref{fig:1}b). 
In the main panel b), the comparison of the experimental dielectric
spectrum~\cite{lunkenheimer2017electromagnetic} with the one calculated 
\cite{holzl2021dielectric} via 
\begin{equation}
    \varepsilon''(\omega) = \frac{\omega}{3\varepsilon_0 k_{\rm B} T V}   \int\limits_0^{\infty} \exp(-i\omega t) \left< \bm M(t) \bm M(0) \right> \text{d}t , 
    \label{eq:eps}
\end{equation}
where $\bm M$ is the total dipole moment of the simulation box, 
demonstrates quantitative agreement between AIMD and experiment
also for this observable;
note that Eq.~(\ref{eq:eps}) also contains  the
harmonic 
quantum correction
factor~\cite{ramirez2004quantum}
consistent with the one applied when computing the Raman spectrum. 

For the Raman spectrum, the negative cross part leads to the fact that the
total correlation function decays on slightly shorter times than the self part.
However, the correlation time of the Raman self part of around \SI{1.2}{ps}
can still not be brought into accordance with the correlation time of the
dielectric self part of approximately \SI{5.3}{ps}, heralding distinctly
different underlying microscopic relaxation mechanisms.

Based on this observation, we now scrutinize the earlier suggestion that
DID effects are dominating the low-frequency Raman spectrum.
In order to disentangle them from the intrinsic polarizability tensor of
the molecules in the liquid state
($\bm \alpha^{\rm eff} = \bm \alpha^{\rm intr,liq} + \bm \alpha^{\rm DID,liq}$),
we use a method introduced earlier~\cite{heaton2006condensed}: 
The total field change at each molecule is given by  
\begin{equation}
    \bm f_{i} = \bm E_0 + \sum\limits_{j\neq i}^N \bm{ T}_{ij}\, \bm \mu_{j}^{\rm ind}
 = \left( \bm 1 + \sum\limits_{j\neq i}^N \bm{ T}_{ij}\, \bm \alpha^{\rm eff}_j \right) \bm E_0,
\end{equation}
where the dipole interaction tensor is given by
$
\bm{ T}_{ij} 
= \left( 3 \bm r_{ij}\bm r_{ij}^T - \bm{1} r_{ij}^2 \right) / r_{ij}^5
$
, and the intrinsic polarizability tensor of molecule $i$ can be obtained from
$
 \bm \alpha_i^{\rm intr,liq} = \bm \alpha^{\rm eff}_i \Big( \bm 1 + \sum_{j\neq i}^N \bm{ T}_{ij}\, \bm \alpha^{\rm eff}_j \Big)^{-1}
$
, since
$\bm \alpha_i^{\rm intr,liq} \bm f_{i} = \bm \alpha^{\rm eff}_i \bm E_0$
must hold for all possible $\bm E_0$.

\begin{figure}[htb!]
    \centering
    \includegraphics[width=1\linewidth]{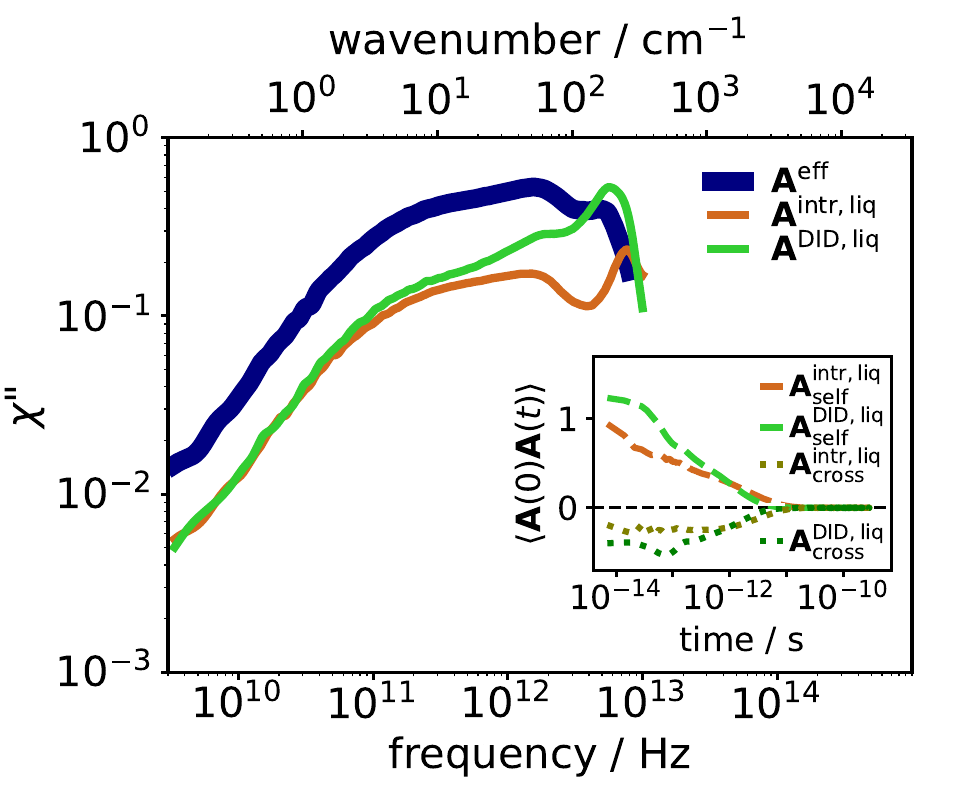}
    \caption{Comparison of the DID and intrinsic contribution to the total Raman
spectrum. Spectra are calculated by inserting the respective polarizability tensor as indicated in the legend into Eq.~(\ref{eq:chi}); see also text. 
Inset: Different contributions to the polarizability correlation functions underlying the spectra in the main panel.}
    \label{fig:2}
\end{figure}

The spectra calculated by employing either the total intrinsic or the total
DID polarizability tensor when evaluating Eq.~(\ref{eq:chi}) are shown
in Fig.~\ref{fig:2}.
Several observations can be made:
First of all, the intensity of the DID contribution is
similar to
that of the intrinsic polarizability at low frequencies.
This clearly shows that
neither of the two can be neglected in this frequency range,
contrary to earlier claims. 
In other words: The DID contribution is not at all negligible in this
frequency range as sometimes assumed in theoretical models~\cite{keyes2024comment}, nor is the rotational contribution, contrary to early force field simulations \cite{frattini1987induced}.
Furthermore, also the shape of the low-frequency mode is basically
identical for both contributions, highlighting their equal share
in shaping the total light scattering spectrum at low frequencies.
The inset of Fig.~\ref{fig:2} shows the self and cross parts of
the two contributions.
Equivalent observations as for the $\bm \alpha^{\rm eff}$ above can be made,
namely that the self parts relax on slightly longer time scales than the total
contribution, while the cross parts are negative. 

Now, one could expect the self part of the
intrinsic polarizability to reflect the single molecule rotational motion,
and thus to be in agreement with the dielectric self part.
The
peak of the lowest mode in the $\bm A_{\rm self}^{\rm intr,liq}$
spectrum is approximately
at \SI{70}{GHz}, 
which is in line with the single-molecule rotational mode in 
the dielectric spectrum at \SI{30}{GHz} after
accounting for a theoretical possible factor up to
three \cite{diezemann1999revisiting}.
In this context, it must be pointed out
that the intrinsic polarizability not
necessarily only reflects the rotational motions.
This is because of internal field effects, i.e., surrounding molecules
altering the electronic configuration of the molecule under consideration,
which render the molecular polarizability tensor different from the one
the molecule would have if placed in vacuum, even for the case of total
absence of DID effects.
Since the surrounding molecules are continuously moving, this alteration
of the electronic structure is a dynamic effect, superimposing the
genuine single-molecule 
rotational motion of the molecule under consideration.

In order to single out the pure rotational contribution, we conceived the
following route:
Each single molecule is taken out of its liquid configuration and placed in
vacuum where, maintaining its exact atomic coordinates, its polarizability
tensor is re-calculated, yielding $\bm \alpha^{\rm intr,vac}$.
In this way, all effects of the surrounding molecules on the molecule under
consideration regarding the electronic structure are eliminated, while preserving
all atomic dynamics along the very same AIMD trajectory. 
Then, the DID contribution can be calculated to first order in molecular
interactions, which has been shown to be sufficient for
water \cite{lupi2012dynamics}, as follows
\begin{equation}
    \bm\alpha_i^{\rm DID,vac} = \sum\limits_{j\neq i}^{N}\bm \alpha^{\rm intr,vac}_i\, 
\bm{ T}_{ij}
\, \bm \alpha^{\rm intr,vac}_j \enspace .
    \label{eq:DIDvac}
\end{equation}
Note that this (approximate) approach is similar, but not equivalent, to the
way Raman spectra have been calculated from 
force field 
molecular dynamics
simulations.
There, however, the polarizability tensor of a water molecule, obtained either
from gas phase experiments \cite{frattini1987induced} or quantum chemical
calculations \cite{lupi2012dynamics}, has been fixed to the atomic frame in the
simulations.
In this way, intramolecular atomic movements are disregarded, which is not the
case in our approach.
Still, it has to be emphasized that also our procedure is not expected to yield
results comparable to experimental data, it is just meant to serve as a rigorous
reference to the full spectrum in the liquid state in order
to probe internal field effects.

\begin{figure}[htb!]
    \centering
    \includegraphics[width=1\linewidth]{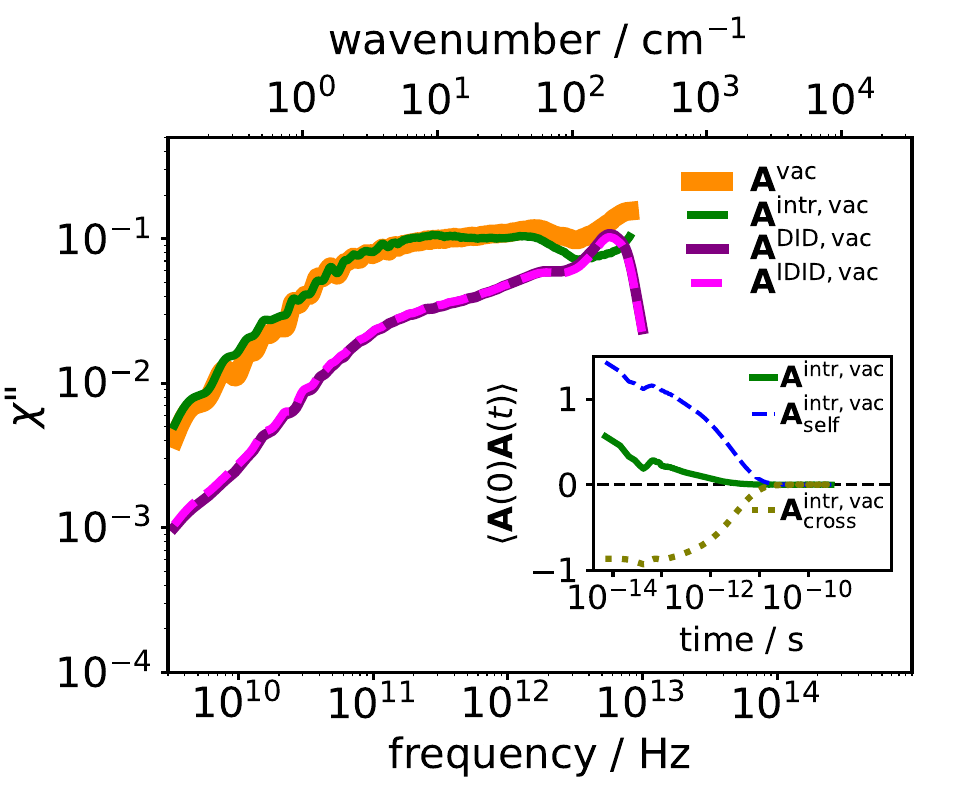}
    \caption{Comparison of the DID, IDID and intrinsic contribution to the total 
Raman spectrum for the calculations "in vacuum", see text. 
Spectra are calculated by inserting the respective polarizability tensor as indicated in the legend into Eq.~(\ref{eq:chi}). 
Inset: Total, self and cross contributions to the intrinsic correlation function. }
    \label{fig:3}
\end{figure}

The resulting spectra are shown in Fig.~\ref{fig:3}, where also the purely
translational contribution $\bm A^{\rm IDID,vac}$ to the DID spectrum is included.
It is calculated \cite{mazzacurati1989low} by replacing the full molecular
polarizability tensors in Eq.~(\ref{eq:DIDvac}) by one third of their trace, 
$\rm{Tr}\, \bm \alpha^{\rm intr,vac}_i/3$, thus excluding the rotational
contributions to $\bm\alpha_i^{\rm DID,vac}$, while keeping only the
isotropic (I) part. 
It can be seen in Fig.~\ref{fig:3} that $\bm A^{\rm DID,vac}$ and
$\bm A^{\rm IDID,vac}$ are basically identical, which quantitatively confirms
the common notion that the DID contribution reflects translational motions. 

When comparing the $\bm A^{\rm DID,vac}$ to $\bm A^{\rm intr,vac}$ spectra,
the following observation
can be made:
The DID contribution is approximately a factor of four weaker in intensity
than the intrinsic contribution at low frequencies, 
highlighting the fact that the influence of the intrinsic field
can markedly influence the weighting of the two contributions.

Now it is interesting to focus on the $\bm A^{\rm intr,vac}$ part, as it does
not contain local field effects, but solely reflects the rotational motions
of the polarizability tensor.
In the inset of Fig.~\ref{fig:3}, the splitting into self- and cross-correlation
contributions is shown, which are both distinctly different from the total
$\left< \bm A^{\rm intr,vac}(0)\bm A^{\rm intr,vac}(t)\right>$, with the
cross-correlation part being strongly negative.
Moreover, the correlation decays on a notably longer timescale in the case of
self and cross part as compared to the total.
In fact, theoretical considerations predict that the time scale of the total
rotational correlation function $\tau_{\rm t}$ is shorter than the one of the
single molecule $\tau_{\rm s}$ in case of negative orientational
cross-correlations~\cite{berne2000dynamic}. 
 
\begin{figure}[htb!]
    \centering
    \includegraphics[width=1\linewidth]{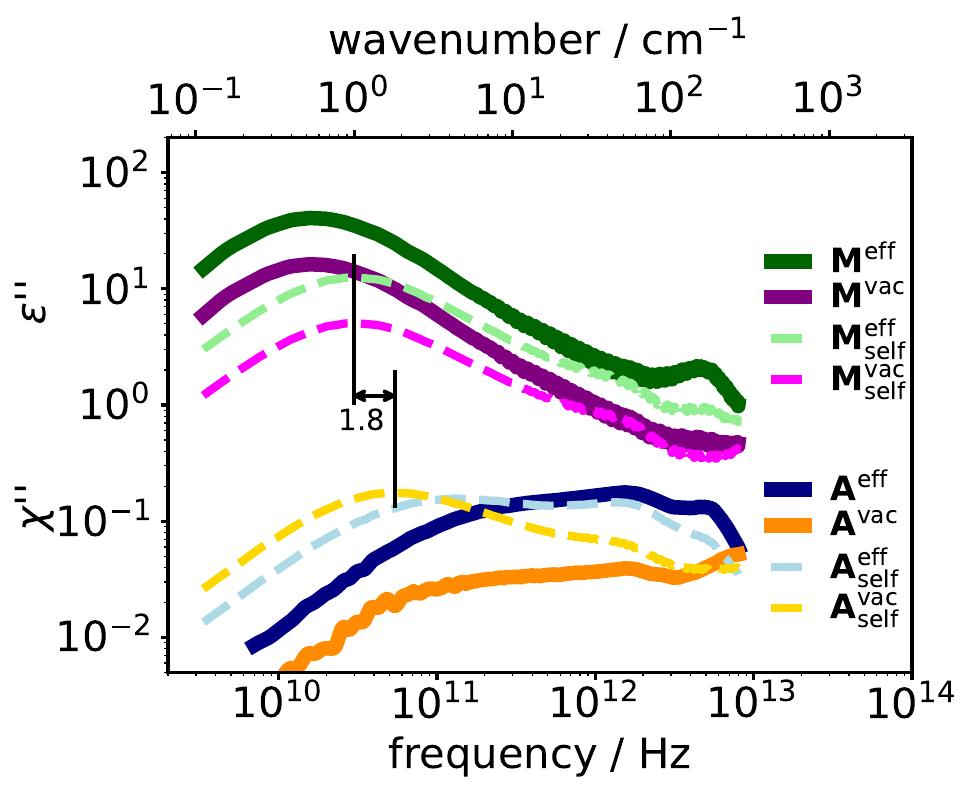}
    \caption{Comparison of dielectric and Raman spectra, calculated by inserting the respective dipole moment into Eq.~(\ref{eq:eps}) or polarizability tensor into Eq.~(\ref{eq:chi})
as indicated in the legend, see text.}
    \label{fig:4}
\end{figure}

After having disentangled the single molecule rotation contribution to the
light scattering spectrum, we can compare it directly to the dielectric spectrum.
To this end, we compute the dielectric loss via Eq.~(\ref{eq:eps})
in two different ways: 
Firstly, in the liquid state \cite{holzl2021dielectric}, i.e., taking all
internal field effects into consideration via the maximally localized
Wannier centers.
The dipole moment calculated in this way is named $\bm M^{\rm eff}$ in analogy
to the corresponding polarizability tensor.
Secondly, the dipole moment "in vaccum" $\bm M^{\rm vac}$ is calculated in
analogy to the polarizability tensor "in vaccum" as explained above.
The comparison of the different dielectric spectra to the corresponding 
Raman spectra is shown in Fig.~\ref{fig:4}.

One can see that the main difference between the dielectric spectra calculated
via these two routes is just the intensity.
This is true for the total as well as the self-correlation part of the spectrum.
Most importantly, the position of the low frequency peak does not change in
going from the liquid calculations to the "in vacuum" ones.
This means that the internal field just alters the apparent magnitude of the
dipole moment, which is well understood \cite{bottcher1974theory}.

In contrast, the situation is distinctly different for the light
scattering spectrum.
Here, the spectra calculated from $\bm A^{\rm eff}$ and $\bm A^{\rm vac}$ are
different not only in intensity but also in shape, due to a different weighting
of the individual contributions.
Especially the position of the low frequency peak in the self part of the two
spectra is quite different.
Notably, it can be seen that the "in vacuum" self part of the Raman spectrum,
which is mainly due to single molecule rotation, is only a factor of 1.8 higher
in frequency than the single molecule rotation in the dielectric spectrum.
This is now in accordance with simple considerations based on the different
ranks of Legendre polynomials involved in the dipole and polarizability
correlation functions.
While the dipole is a vectorial quantity, thus corresponding to a first rank
correlation function, the polarizability tensor is second rank.
Therefore, a factor of three between the peak positions of the two methods can
be expected if rotational diffusion is the molecular mechanism underlying the
dynamics and the same peak position is expected for large angle
jumps \cite{diezemann1999revisiting}.
Here, the observed factor 1.8 versus 3 implies significant deviations from
rotational diffusion, i.e., small to medium size angular jumps, in good
agreement with the current picture of H-bond dynamics in water 
in terms of the molecular jump mechanism of water reorientation \cite{%
laage2006molecular,laage2011anrev}.

In summary, supported by the excellent agreement of the computed total low-frequency
Raman spectrum compared to experiment, we decompose it into various
contributions which unfold an astonishingly complex behavior, 
including distinct differences from dielectric loss spectra. 
We show that translational motions, contributing via the dipole-induced-dipole effect,
are equally important as the contribution from the intrinsic
polarizability, which contains rotational motions and internal
field effects. 
The pure total rotational contribution, on the other hand, is shifted to higher
frequencies as compared to the isolated single molecule rotation, due to strong
negative orientational cross-correlations in conjunction with internal field effects.
The latter alter the electronic structure of molecules embedded in the liquid 
as a result of electronic polarization and charge transfer effects compared to
isolated molecules. 
The comparison of the Raman to the dielectric spectrum reveals a rotational jump
mechanism as well as that the internal field effect on dielectric loss is much
simpler than on Raman response.
Overall, our study provides a comprehensive explanation for the microscopic
origin of water's lowest-frequency Raman mode.
Having understood the pure solvent, we expect our molecular insights to be
highly relevant for correctly interpreting the Raman spectra of complex aqueous
solutions, as customarily used to investigate how the dynamics of water
molecules are influenced by the presence of ions or
biomolecules~\cite{perticaroli2011extended,lupi2012dynamics,gonzalez2023lifting,zeissler2025fresh}.

\section*{Acknowledgments}
We are grateful to Catalin Gainaru for providing the depolarized light scattering data from Ref.~\cite{hansen2016identification}.
%
%
Funded by the Deutsche Forschungsgemeinschaft
(DFG, German Research Foundation) under Germany's
Excellence Strategy~-- EXC~2033~-- 390677874~-- RESOLV.
%
The authors gratefully acknowledge the computing time made available to them on the high-performance computer Noctua~2 at the NHR Center Paderborn Center for Parallel Computing (PC2). This center is jointly supported by the Federal Ministry of Research, Technology and Space and the state governments participating in the National High-Performance Computing (NHR) joint funding program (www.nhr-verein.de/en/our-partners).

\bibliography{bib-v6}

\end{document}